\documentclass{jps-cp}
\usepackage{txfonts} 

\newcommand{\baryon}{{\rm b}}

\newcommand{\obh}{$\Omega_{\baryon}{\cdot}h^2$}

\newcommand{\dpg}{{D(p,$\gamma)^3$He}}

\newcommand{\hli}{{$^4$He, D, $^3$He and $^{7}$Li}}
\newcommand{\sfac}{$S$--factor}
\newcommand{\zaa}{Astron.~Astrophys.}

\newcommand{\zapj}{Astrophys.~J.}
\newcommand{\zapjl}{Astrophys.~J.~Lett.}

\newcommand{\znat}{Nature}

\newcommand{\znp}{Nucl.~Phys.}

\newcommand{\zpr}{Phys.~Rev.}

\newcommand{\zprl}{Phys.~Rev.~Lett.}

\newcommand{\zadndt}{At. Data Nucl. Data Tables}

\newcommand{\zjcap}{J. Cosmol. Astropart. Phys.}

\title{Precision Big Bang Nucleosynthesis with the New Code {\it PRIMAT}}

\author{Cyril  \textsc{Pitrou}$^{1,2}$, Alain  \textsc{Coc}$^3$,  Jean--Philippe  \textsc{Uzan}$^{1,2}$ and Elisabeth  \textsc{Vangioni}$^{1,2}$}

\inst{${}^1$Institut d'Astrophysique de Paris, CNRS UMR 7095, 98bis Bd. Arago, 75014 Paris, France\\
${}^2$ Sorbonne Universit\'e, Institut Lagrange de Paris, 98bis Bd. Arago, 75014 Paris, France\\
${}^3$Centre de Sciences Nucl\'eaires et de Sciences de la Mati\`ere, CNRS, B\^at. 104, F-91405 Orsay Campus, France}

\email{coc@csnsm.in2p3.fr}

\recdate{July 31, 2019}

\abst{
Precision on primordial abundances, deduced from observations, have now reached 
the percent level for $^4$He  and deuterium. Precision on  big bang nucleosynthesis (BBN) predictions should, 
hence, reach the same level. The uncertainty on the $^4$He  mass fraction is strongly affected by 
theoretical uncertainties on the weak reaction rates that interconvert neutrons with protons. 
All these corrections have been calculated in a self-consistent manner and implemented in
a new, and public,  {\it Mathematica} code {\it PRIMAT} \cite{Pit18}, 
together with an extensive data base of reaction rates. Both can be obtained at {\tt http://www2.iap.fr/users/pitrou/primat.htm}.

}

\kword{Big bang nucleosynthesis, Helium--4, Deuterium, Mathematica notebook}

\begin{document}
\maketitle

\section{Introduction}

Precision on primordial abundances, deduced from observations, have now reached 
the percent level for $^4$He (1.6\% \cite{Ave15}) and D/H  (1.2\% \cite{Coo18}) . 
The baryonic density of the Universe is now known at better than the percent level 
(0.7\%), thanks to the observations of the anisotropies of the
Cosmic Microwave Background radiation by the {\it Planck} mission \cite{Planck16,Planck18}.  
Precision on BBN predictions should, hence, reach the same level (see e.g.  Cyburt et al. \cite{Cyb16} for a review), 
but is mostly limited by the precision of experimental nuclear cross sections. 
On the contrary, even though the uncertainty on the $^4$He  mass fraction comes 
partially from the experimental  neutron lifetime, it is also strongly affected by 
theoretical uncertainties of the weak reaction rates that interconvert neutrons with protons. 
Those weak rates are mainly influenced by radiative corrections and by finite nucleon mass effects. 
In addition, the thermodynamics in the early Universe is affected by the incomplete decoupling of 
neutrinos and QED plasma effects. 
For all of these corrections, detailed balance was enforced, an essential requirement as there 
rates govern the n/p ratio at freeze-out and, consequently the $^4$He  abundance. 
Precision on D/H is also important for the {\em lithium problem}: the predicted lithium 
abundance is a factor of three higher than observations.
As most proposed solutions result in an increase of D/H, not compatible with observations,
precision on D/H is required.  

\section{The {\it PRIMAT} code}

We had developed a {\it Fortran77} code ({\it EZ\_BBN}) \cite{CV17} together with a network of reactions,
including n, p, d, t, $^3$He and $\alpha$ captures on nuclei up to the CNO region,
with the most up to date nuclear input \cite{Coc12a,Coc15,Ili16,Gom17}. It has been adapted to 
investigate, beyond the standard model physics, like modified gravity \cite{COUV06,COUV08}, 
variation of fundamental constants \cite{CNOUV06}, or exotic neutron sources \cite{neutrons,miroir1,miroir2}. 
We have recently developed a new code, {\it PRIMAT} (for PRImordial MATter), \cite{Pit18} in
a {\it Mathematica} platform to more easily incorporate the complex algebra required to calculate
the corrections to the weak rates and thermodynamics. Those were only partially implemented in the
previous code. A comparison between the results of the two codes are displayed in Table~\ref{t:abund}.
The two codes share the same network and are in very good agreement, except for $^4$He because of the 
improved weak rate calculations in {\it PRIMAT}. This gives confidence in the results since the two codes
were developed, almost independently.

\begin{table}
\caption{Predictions compared to observations.}
\begin{center}
\begin{tabular}{lcccc}
\hline
& $Y_p$ & D/H$\times10^5$&  $^3$He/H$\times10^5$ & $^7$Li/H$\times10^{10}$\\
\hline
Observations & 0.2449$\pm$0.0040 \cite{Ave15} & 2.527$\pm$0.030 \cite{Coo18} & $<$(0.9-1.3) \cite{Ban02} & 1.58$\pm$0.31 \cite{Sbo10} \\
\hline
{\it EZ\_BBN} \cite{Coc15} &  0.2484$\pm$0.0002  &2.45$\pm$0.05 & 1.07$\pm$0.03 &5.61$\pm$0.26 \\
{\it PRIMAT} \cite{Pit18} &0.24709$\pm$0.00017 &  2.459$\pm$0.036 & 1.074$\pm$0.026 & 5.623$\pm$0.247 \\
\hline
\end{tabular}
\label{t:abund}
\end{center}
\end{table}

Even if they are small, the corrections affect the predictions by 1.84\% for $^4$He  
and 1.49\% for deuterium, larger that the observational uncertainties. Hence, they are 
essential, not only for the $^4$He  abundance prediction but also for deuterium,
as shown in Table~\ref{t:cor}. Among the many corrections discussed in detail
in \cite{Pit18} (and references therein), the most important for the $^4$He mass fraction ($Y_p$), are the radiative corrections and the finite mass 
of the nucleon (i.e. the c.m. frame is not exactly the nucleon frame). 
For D/H, it is also important (Table~\ref{t:cor}) to take into account the incomplete neutrino decoupling. In our
previous code, we assumed that neutrinos decouple from the plasma at $T\approx$2--3~MeV \cite{Dol02}, 
well before electron positron--annihilation, so that they do not take advantage of the re-heating process
that affects the other components of the plasma. We now use a parametrization \cite{Pis08} to take  this effect into account.

\begin{table}[h]
\caption{Effect of main corrections on $Y_P$ and D/H.}
\begin{center}
\begin{tabular}{lcccc}
\hline
& ${\Delta}Y_p\times10^4$ & ${\Delta}Y_p$ (\%)&   ~~~ & $\Delta(D/H)$ (\%)\\
\hline
Radiative corrections & 31 &1.23 && 0.70 \\
Finite mass of nucleons & 13 & 0.53 && 0.25 \\
Incomplete neutrino decoupling &  &&& 0.37 \\
QED effects on plasma & &&& 0.12 \\
\hline
Total & 44 & 1.84 && 1.44 \\
\hline
\end{tabular}
\label{t:cor}
\end{center}
\end{table}

A set of files can be obtained from  {\tt http://www2.iap.fr/users/pitrou/primat.htm}, allowing one to re-do
all calculations presented in the Pitrou et al. \cite{Pit18} paper. Among these files, we emphasize the following:

\begin{itemize}
\item {\tt PhysReptArxivVersion.pdf}: an updated version of the original publication \cite{Pit18}, including the corrections
of a few typos (in the paper not in the code). See e.g. Eqs. (62) and (B23) and RC+FM+WM+ID line in Table V.
\item {\tt PRIMAT-Main.nb} the main Mathematica "notebook" (and its pdf version, {\tt PRIMAT-Main.pdf}, for examination only).
\item {\tt BBNRatesAC2019.dat}, the tabulated reaction rates (see \S~\ref{s:rates}). 
\item {\tt PRIMAT\_small\_network.nb}, a small 12+1 reaction network using 
{\tt BBNRates\_2018\_12reactions.dat} for the 12 tabulated rates.
\item {\tt PRIMAT-MonteCarlo-Rates1.nb} to calculate the primordial abundances with Monte Carlo calculated errors 
bars, for a single \obh$\pm\Delta$ \obh.
\item {\tt PRIMAT-Abundances-Eta1.nb}  is the same but produce the nice \hli\ versus $\eta$ (or \obh), 
"Schramm" plot.
\item {\tt PRIMAT-Abundances-and-Plots1.nb} produces a nice plot of  A=1 to 23 abundances as functions of time.
\end{itemize}

\section{Reaction rates}
\label{s:rates}

The {\it PRIMAT} package includes, besides a few reaction rates represented by analytical expressions within the code,
a file where the other reaction rates are given in tabular format: {\tt BBNRatesAC2019.dat}. 
Each carries also its reverse ratio so that the reverse reaction is also systematically included in the network. 
The file can be easily edited to accommodate alternative rates or
even new reactions. It can also be used as a benchmark to compare different codes as most differences in
BBN results come from the use of different reaction rates. 
The file was finalized at the end of 2017 and is expected to be updated by the end of 2019,
in particular to take into account the following evolutions.        

\begin{itemize}
\item The \dpg\ reaction is crucial for the calculation of the deuterium abundance as theoretical predictions of its
rate differ significantly \cite{Mar05,Mar16}, without sufficient experimental data to discriminate between them.
This is expected to evolve with the forthcoming release of experimental data obtained at LUNA \cite{LUNA}.  
\item The $^7$Be(n,p)$^7$Li rate is currently being re-evaluated \cite{THK} to take into account new measurements \cite{Dam18,Tom19,proc1,proc2}.
\item The $^7$Be(n,$\alpha$)$^4$He. Because of the identical nature of the two outgoing $\alpha$--particles and parity conservation,
this rate is much lower than that of $^7$Be(n,p)$^7$Li. Thanks to recent experiments \cite{Hou15,Bar16,proc1} this
has been confirmed.
\item The  $^7$Be+d reaction deserves peculiar attention. In a recent experiment \cite{Rij19},
a new resonance was found in the, previously unexplored, $^7$Be(d,$\alpha$)$^5$Li(p$\alpha$) channel. 
Accordingly, their experimental \sfac\ is significantly higher\cite{Rij19}  than the previous experimental 
ones \cite{Kav60,Ang05}, that overlooked this channel. 
However, because of the the "safety" factor introduced by Parker  \cite{Par72} "to take into account 
contributions from higher excited states in $^8$Be.'', by pure chance, the recommended rate is very close to the commonly 
adopted rate \cite{CF88} in BBN calculations.
This is, indeed, the rate  \cite{CF88} that we have been using so far, in particular in {\it PRIMAT}, together with an
estimated factor of three uncertainty. Besides, it has long be known \cite{Coc04,Coc12a,Kir11} 
that an increase of this rate by factor of $>$30 is required to have a significant effect on $^7$Be/$^7$Li.
Hence, this new measurement \cite{Rij19} is welcome as it puts the $^7$Be+d rate and uncertainty on firmer ground but
{\em it also confirms that this reaction rate is too small to affect BBN, even at the percent level (contrary to Ref.~\cite{Rij19} conclusions). }    
\end{itemize}

\end{document}